\documentclass{article}
\usepackage{frascatiphys,here,graphicx,subfigure}
\usepackage{xspace}
\newcommand \B {\ensuremath B\xspace}
\newcommand \D {\ensuremath D\xspace}
\newcommand \susy {\rm SUSY\xspace}
\newcommand \CP {\ensuremath CP\xspace}
\newcommand \tev {\rm TeV\xspace}
\newcommand \gev {\rm GeV\xspace}

\newcommand \e[1] {\ensuremath 10^{#1}}
\def \stwobeff {\ensuremath \sin(2 \beta_{\rm eff})}
\def\iab{\ensuremath ab^{-1}}
\begin{document}
\title{
THE PHYSICS CASE OF THE SUPERB FACILITY
}
\author{
A.~Bevan       \\
{\em Particle Physics Research Center, Queen Mary University of London,}\\
{\em Mile End Road, London, E1 4NS, United Kingdom.} \\
}
\maketitle
\baselineskip=11.6pt
\begin{abstract}

The physics case of the SuperB facility with design luminosity of $10^{36}$ ${\rm cm}^{-2}{\rm s^{-1}}$
is compelling.  Such a facility has a rich and varied potential to probe physics
beyond the Standard Model.  These new physics constraints are obtained through the
study of the rare or Standard Model forbidden decays of $B_{u,d,s}$, $D$ and $\tau$ particles.
The highlights of this wide-ranging physics programme are discussed in these proceedings.
\end{abstract}
\baselineskip=12pt
\section{Introduction}
A conceptual design report of a next generation $e^+e^-$ collider capable of delivering 100 times the
luminosity of the current \B factories has recently been
compiled~\cite{bevan:superbcdr}.  This report forms the basis of the physics motivation,
detector, and accelerator designs for the next generation \B factory at an $e^+e^-$ collider.
Details of the accelerator
and detector designs are discussed elsewhere~\cite{bevan:superbintheseprocs}.
Data taking could commence as early as 2015 if the project is approved in the next few years.
By this time, the LHC will have
produced the results of direct searches for \susy, Higgs particles and many other new
physics (NP) scenarios, as well as providing precision measurements of \CP violation and the
CKM mechanism for quark mixing in $B_{u,d,s}$ decays.
  The focus of high energy physics at that time
will either be to understand the nature of any new particles found at the LHC, or to try and
indirectly constrain possible high energy new particles by looking for virtual contributions
to increasingly rare decays.  If new particles exist they can contribute significantly
at loop level to many rare $B$, $D$ and $\tau$ decays.  If this occurs, we may measure observables that differ from
Standard Model (SM) expectations.  Precision measurements
of branching fractions, \CP, and other asymmetries in many different rare decays can be
used to elucidate the flavor structure of new particles and distinguish between different
NP scenarios.  Some NP scenarios introduce new particles at low energies
(few \gev) which can be observed directly at SuperB. In short, the main aim of the SuperB facility
is to search for and elucidate the behavior of NP.

\section{\B Phyiscs}
\subsection{Measurements of $\stwobeff$}
\begin{figure}[H]
    \begin{center}
        {\includegraphics[scale=0.29]{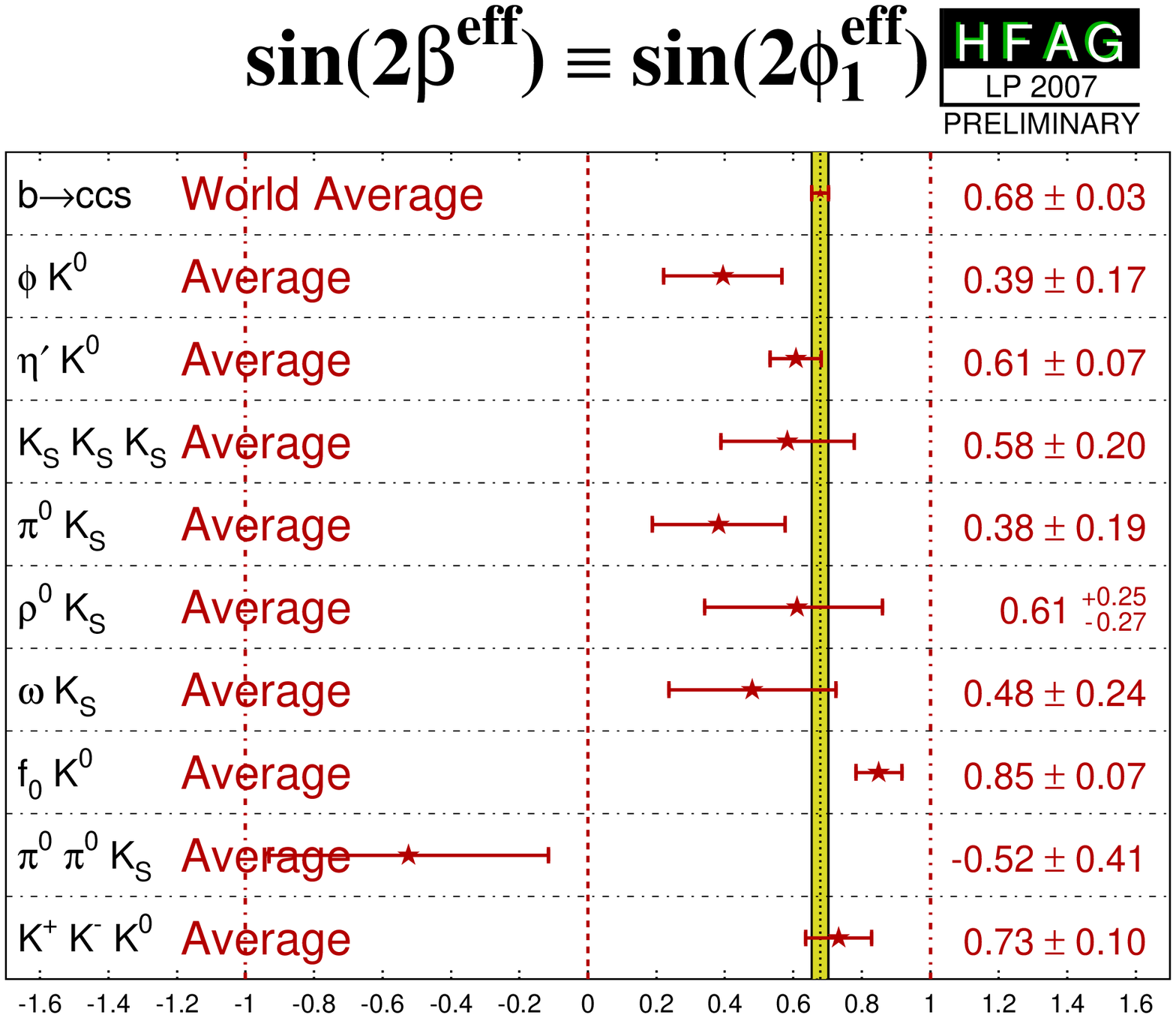}}
        \caption{\it The distribution of $\stwobeff$ measured in $b \to s$ penguin decays, along with the reference
        measurement of $\sin(2\beta)$ from $b \to c\overline{c}s$ decays}
\label{fig:hfag_sin2beta}
    \end{center}
\end{figure}

Since the discovery of \CP\ violation in the decay of B mesons through $b\to c\overline{c}s$ transitions,
an industry has developed in performing alternate measurements of $\sin 2\beta$ in other processes
(these are measurements of $\stwobeff$).  In the presence
of new physics, one can measure \CP\ asymmetries that are significantly different from
the SM expectation which is $\sin 2\beta$ measured in $b\to c\overline{c}s$ transitions.  Measurements of $\stwobeff$ are performed in decays with $b\to s$ and $b \to d$
transitions. Loop dominated rare decays can receive significant contributions from new physics, and large
effects have been ruled out by current measurements (See Figure~\ref{fig:hfag_sin2beta}).  The general trend of
measurements shows that $\stwobeff < \sin(2 \beta)$.  In addition to the experimental uncertainties on the
measurement of $\stwobeff$, there are theoretical uncertainties on the SM prediction in each decay mode.
The first thing to note when considering theoretical uncertainties is that different decay modes have
different expected shifts that are known with different levels of precision.  As a result, it is not
correct to average all of the $\stwobeff$ measurements and compare this average with the reference from
$b\to c\overline{c}s$ transitions, although in practice this comparison is often made.  You really have to perform a precision measurement for each mode,
and then make the comparison.  Ideally when you make such a comparison, you want the experimental and theoretical
uncertainties to be similar. There is insufficient statistics available at the existing \B factories
to do this comparison correctly.  The most precise estimates of the SM uncertainty on
$\Delta S = \stwobeff-\sin(2 \beta)$ are of the order of a percent for $\B \to \eta^\prime K^0$,
$\B \to K^+K^-K^0$, and $\B \to 3K^0_S$ decays~\cite{bevan:deltastheory}.  SuperB will be able to
experimentally measure $\stwobeff$ to one percent with $75\iab$, thus enabling a comparison
at the few percent level between $\stwobeff$ and $\sin(2 \beta)$.

\subsection{New Physics in Mixing}

\begin{figure}[H]
    \begin{center}
\subfigure[]
        {\includegraphics[scale=0.24]{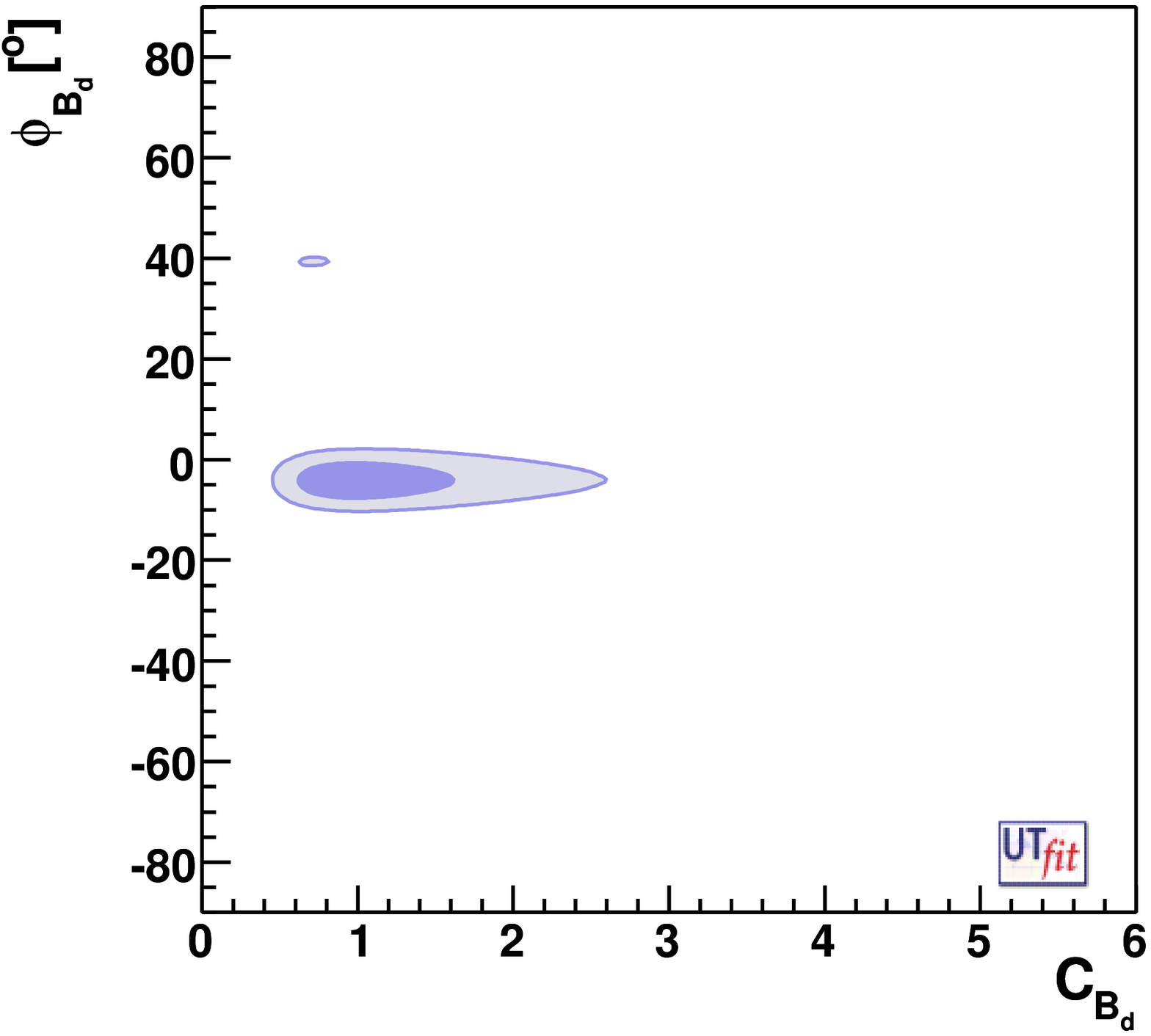}}
\subfigure[]
        {\includegraphics[scale=0.24]{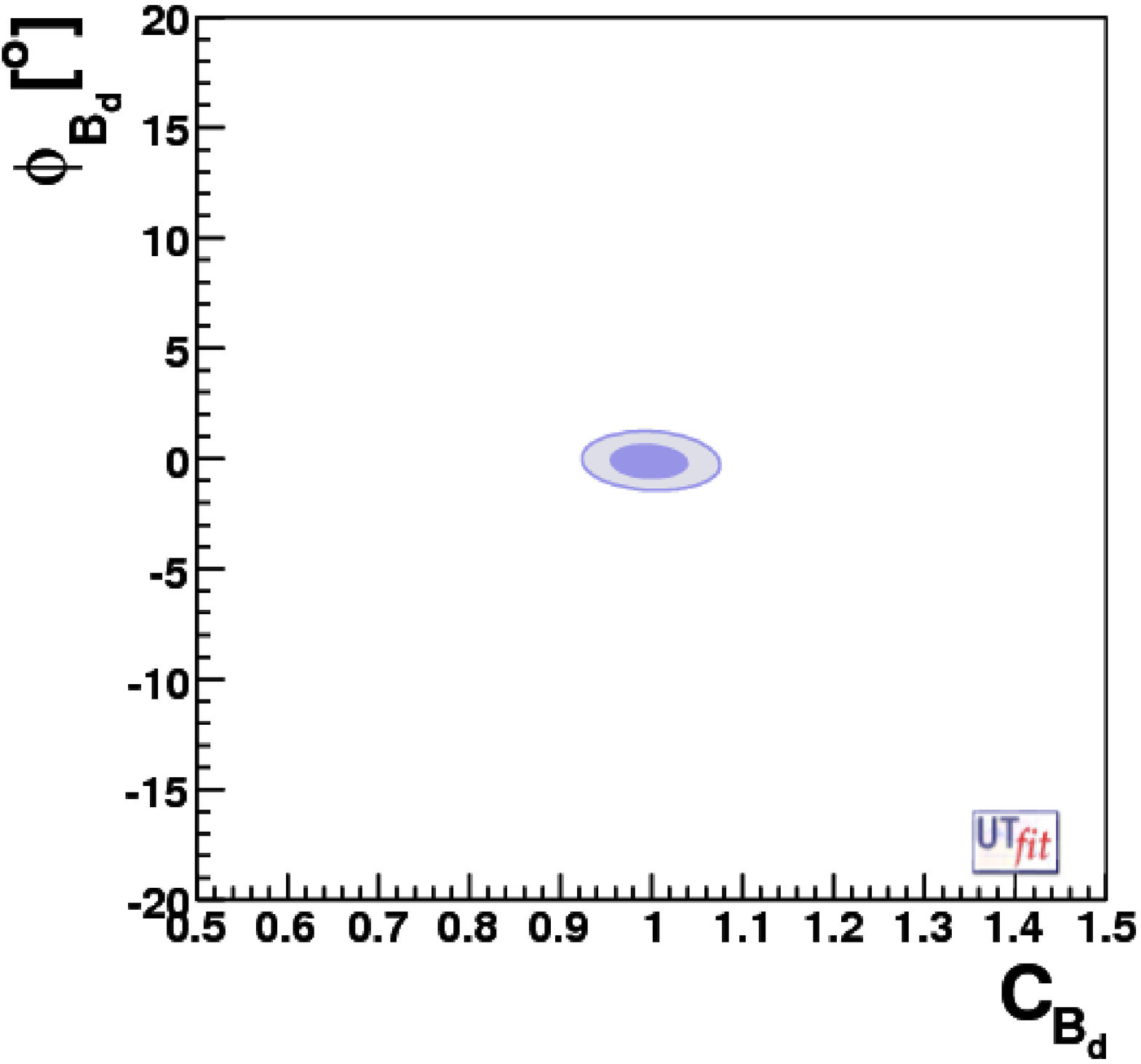}}
        \caption{\it The distribution of $C_{B_d}=1$ vs $\phi_{B_d} =0$ obtained a)
         from current measurements, and b) obtained with 75$\iab$ of data from a SuperB factory.}
\label{fig:npinmixing}
    \end{center}
\end{figure}

In the SM we know that $B_{d}$ and $B_{s}$ mesons mix.  It is possible to model new physics in mixing
by allowing for an arbitrary NP amplitude to also contribute to the box diagram, and search for
the effect of NP by comparing data to the ratio of the NP+SM contribution to that of the
SM~\cite{bevan:npinmixing}, i.e.
\begin{eqnarray}
C_{B_d}e^{i\phi_{B_d}} = \frac{<B^0|{\cal H}_{NP+SM}|\overline{B}^0>}{<B^0|{\cal H}_{SM}|\overline{B}^0>}.
\end{eqnarray}
The SM prediction is for $C_{B_d}=1$ and $\phi_{B_d} =0$, so any deviation from this would signify NP.
It is possible to constrain $C_{B_d}$ and $\phi_{B_d}$ using the available data, and extrapolate to
SuperB as shown in
Figure~\ref{fig:npinmixing}.

\subsection{Minimal Flavor Violation}

\begin{figure}[H]
    \begin{center}
\subfigure[]
        {\includegraphics[scale=0.24]{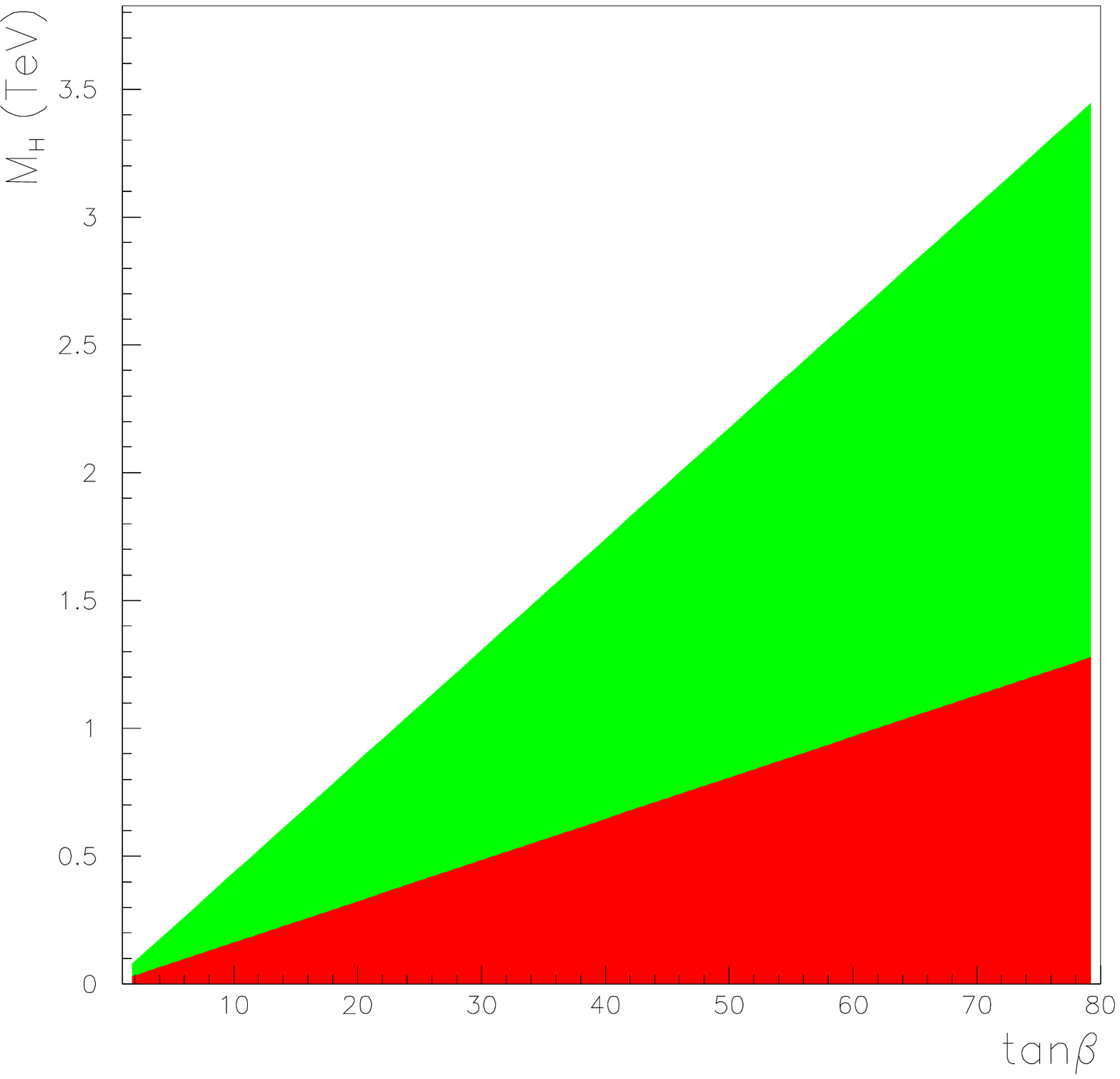}}
\subfigure[]
        {\includegraphics[scale=0.24]{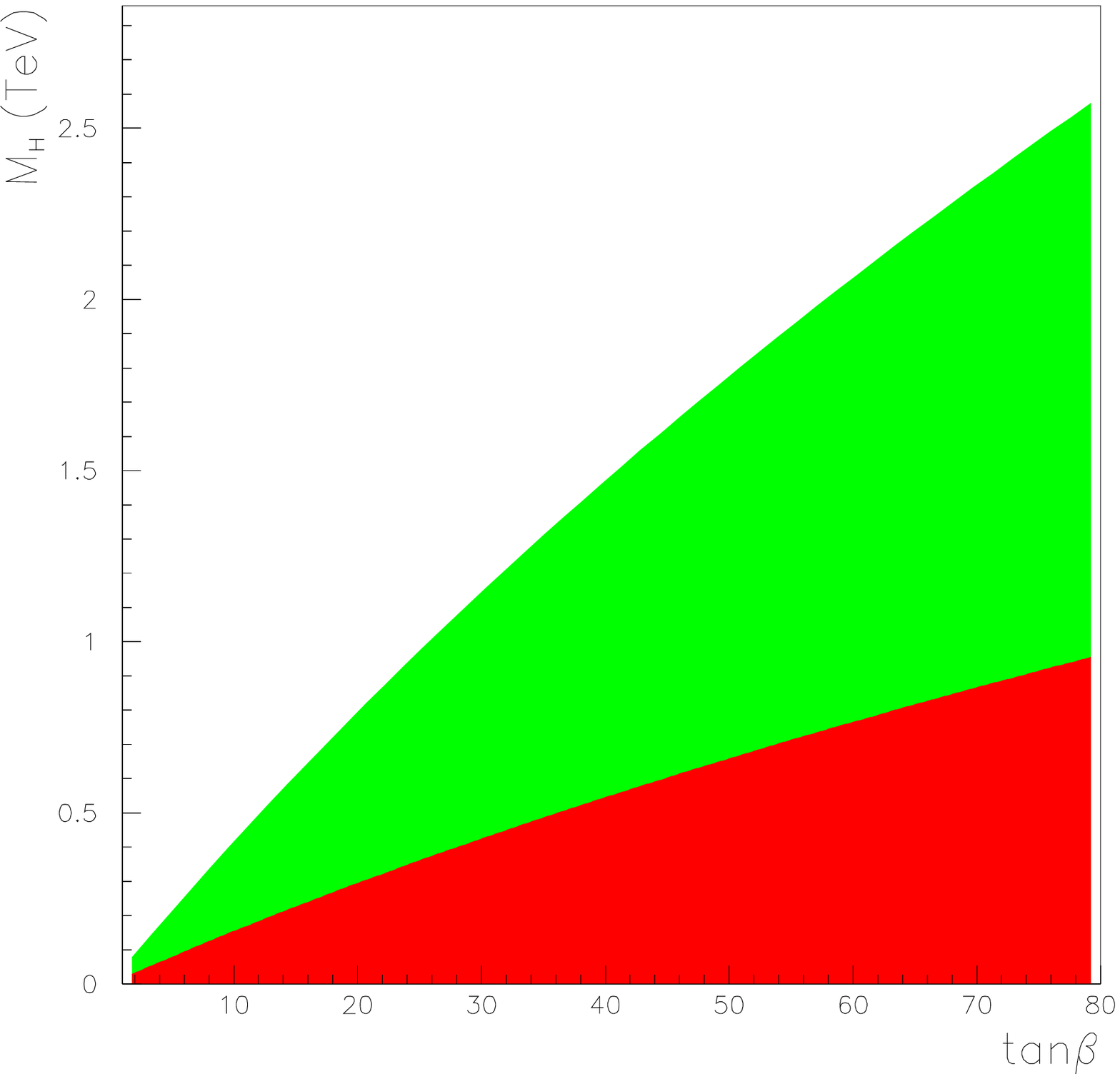}}
        \caption{\it The distribution the mass of the charged Higgs vs $\tan \beta$ in a) 2HDM and
        b) MSSM. The red band is what
        could be excluded by the current \B factories with a data sample of 2$\iab$, and the green
        band is what could be excluded using 75$\iab$ of data from a SuperB factory assuming that the
        measured $B^+ \to \tau^+ \nu$ branching fraction has the standard model value.}
\label{fig:mfv}
    \end{center}
\end{figure}

One set of NP models that is popular assumes that there are no new flavor couplings.  The
corollary of this is that all \CP\ violation is described by the SM Yukawa couplings.
Models of this type are called Minimal Flavor Violation (MFV) models, examples of these are
Higgs doublet, MSSM and large extra dimension models.  Within the realm of MFV models we can
still use the SuperB experiment to tell us about the nature of NP.  For example, it is possible
to use $B^+ \to \tau^+ \nu$ decays to constrain the mass of the
charged Higgs $m_{H^+}$ as a function of the Higgs vacuum expectation value, $\tan \beta$ in
2HDM or MSSM (See Figure~\ref{fig:mfv}).  In 2HDM, the branching fraction
of $B^+ \to \tau^+ \nu$ can be enhanced or suppressed by a factor $r_H$ which has the form
$(1 - \tan^2 \beta[m_B^2/m_{H^+}^2])^2$~\cite{bevan:taunu2hdm}, and the corresponding factor
for MSSM is $(1 - \tan^2 \beta[m_B^2/m_{H^+}^2]/[1+\epsilon_0 \tan \beta])^2$ where $\epsilon_0\sim 0.01$~\cite{bevan:taunumssm}.
Other decay modes, including
$D^+ \to \tau^+ \nu$, $\mu^+ \nu$, and $b \to s \gamma$ can be used to constrain the charged
Higgs mass in a similar way.  The worst case scenario of MFV suggests that SuperB would
be sensitive to new particles with masses up to 600 \gev.

\subsection{Other Searches for New Physics}

\begin{figure}[H]
    \begin{center}
        {\includegraphics[scale=0.29]{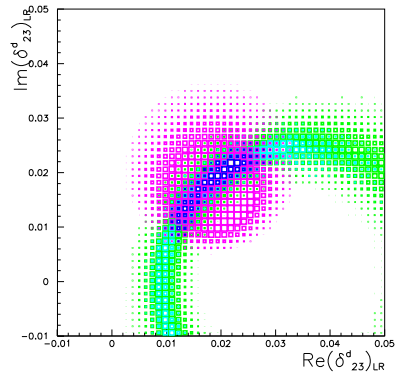}}
        \caption{\it The distribution constraint on the real vs imaginary part of $(\delta^d_{2,3})_{LR}$
        obtainable at SuperB using the constraints described in the text.
        }
\label{fig:squarkmixing}
    \end{center}
\end{figure}

In contrast to the MFV scenario described above, we can think of a more
generalized SUSY scenario. Given that quarks and neutrinos can change type
or mix, it is natural to consider that their super-partners would also
have non-trivial flavor couplings and would mix.  If this is not true, then
the NP extension to the SM would have a fine-tuned and unnatural behavior.
We can already rule out large new physics contributions to \B and kaon physics,
but \CP\ violation is small in the SM, so we should not expect to see
large ${\cal O}(1)$ NP effects, and should be content to search
for small \CP\ violation effects from NP.  The simplest model of this
type is MSSM with squark mixing matrices.  Combinations of observables
measurable at SuperB can be combined to provide non-trivial constraints on
the real and imaginary parts of these squark mixing parameters.  For example,
Figure~\ref{fig:squarkmixing} shows the constraint that SuperB
can put on the complex parameter $(\delta^d_{2,3})_{LR}$ with a data sample of 75$\iab$, where the
$d$ indicates a quark, the indices $2,3$ indicate mixing between
the second and third squark generations and the $LR$ indicates a left-right helicity for the
SUSY partner quarks.  The measurements of the branching fractions of $b\to s \gamma$ (green)
and $b\to s l^+l^-$ (cyan), with the \CP asymmetry in $b\to s \gamma$ (magenta) are combined (blue)
to constrain the real and imaginary parts of $(\delta^d_{2,3})_{LR}$.
SuperB has a sensitivity $>100$ \tev\ for this type of NP model~\cite{bevan:superbiv}.
Other examples of constraints squark mixing parameters are described in~\cite{bevan:superbcdr}.

There are also models of NP that predict light new particles (Higgs or dark matter candidates).  If such
particles exist, then it would be possible to create them directly at a SuperB factory.
Some of these models are described further in~\cite{bevan:superbcdr}.

\section{\D Phyiscs}
Given the recent observation that $D^0$ mesons mix, we now know that the plethora of observables
that one can use to search for \CP\ violation in \B\ decays also exists in \D\ decays.  As with
\B\ meson decays, the pattern of observables (the branching fractions, \CP\ asymmetries and other
observables) in the decays of charm decays can be used to constrain NP scenarios.
Work is ongoing to understand how to use these correlations in charm decays to constrain NP.

\section{$\tau$ Phyiscs}
Many NP scenarios have
couplings that represent lepton flavor violation (LFV).  Such a decay would give an unmistakable signal
in the detector, and would mark the start of a new era in particle physics.  The current best limits from searches
for signals of LFV are ${\cal O}(\e{-7})$~\cite{bevan:taulfv}.  These limits are an order of magnitude away
from upper bounds in many new physics scenarios~\cite{bevan:taulfvnewphysics}.  A SuperB facility would
provide sufficient statistics to find LFV at the level of such predictions, or push upper limits down
to ${\cal O}(\e{-9}$ to $\e{-10})$.

The decays of $\tau$ leptons proceed via a single amplitude.  If a non-zero \CP\ asymmetry is measured
in any $\tau$ decay, then this is a clear signal of new physics.  There have been many proposed searches for
\CP\ violation in $\tau$ decays~\cite{bevan:taucpv}.  When doing such
searches, one has to decouple the possible effects of \CP violation in any final state kaons, and the
difference between $K^+$ and $K^-$ interactions in the detector.

It is possible to test CPT by comparing the ratio of lifetimes of the $\tau^+$ and $\tau^-$.  Any
deviation from one would indicate CPT violation.  The expected statistical precision of such a test
is at the level of ${\cal O}(\e{-4})$.  If this precision were to be achieved, then the lifetime
ratio test in $\tau$ decays would be comparable to that in $\mu$ decays~\cite{bevan:pdg}.

\section{Conclusion}
The SuperB facility has the potential to indirectly search for NP at energy scales far beyond the
reach of the direct searches at the LHC.  The ability to probe flavor
couplings in NP scenarios up to several hundred \tev\ means that results from
SuperB will be of general interest, and complimentary to the LHC physics programme over the
next few decades.  There are two possible scenarios:
(i) If the LHC discovers new particles, it will be possible to measure
their basic properties such as mass and width at ATLAS and CMS.  However, in order to
fully understand these new particles, one needs to understand their
flavor dynamics as well.  The flavor dynamics of new physics can be probed
well above the \tev\ scale at SuperB.  (ii) If the LHC doesn't discover
any new particles, then it is important to probe ever increasing
energy scales.  Again, SuperB can probe well above the \tev\ scale
while indirectly searching for new physics.  The correlations of flavor related observables
measured at SuperB can help us distinguish between the multitude of
NP scenarios being proposed today.  Without this set of
measurements from SuperB, we may not be able to resolve between many of the
plausible NP scenarios that exist.
These proceedings discuss the core of the physics programme of SuperB,
and the interested reader will find a more comprehensive treatment in Ref.~\cite{bevan:superbcdr}.
More discussion on exploiting correlations between measurements of flavor observables
to distinguish between NP models can be found in Ref~\cite{bevan:1036}.
\end{document}